\documentclass[twocolumn,showpacs,preprintnumbers,amsmath,amssymb]{revtex4}

\usepackage{graphicx}
\usepackage{dcolumn}
\usepackage{bm}

\begin{document}

\preprint{APS/123-QED}

\title{Photonic crystals with topological defects}

\author{Seng Fatt Liew}
\affiliation{Applied Physics Department, Yale University, New Haven CT ,USA.}
\author{Sebastian Knitter}
\affiliation{Applied Physics Department, Yale University, New Haven CT ,USA.}
\author{Wen Xiong}
\affiliation{Applied Physics Department, Yale University, New Haven CT ,USA.}
\author{Hui Cao}
\email{hui.cao@yale.edu}
\affiliation{Applied Physics Department, Yale University, New Haven CT ,USA.}
\date{\today}

\begin{abstract}
We introduce topological defect to a square lattice of elliptical cylinders. 
Despite the broken translational symmetry, the long-range positional order of the cylinders leads to residual photonic bandgap in the density of optical states. 
However, the band-edge modes are strongly modified by the spatial variation of ellipse orientation. 
The $\Gamma-X$ band-edge mode splits into four regions of high intensity and the output flux becomes asymmetric due to the formation of crystalline domains with different orientation. 
The $\Gamma-M$ band-edge mode has the energy flux circulates around the topological defect center, creating an optical vortex. 
By removing the elliptical cylinders at the center, we create localized defect states which are dominated by either clockwise or counter-clockwise circulating waves. 
The flow direction can be switched by changing the ellipse orientation. 
The deterministic aperiodic variation of the unit cell orientation adds another dimension to the control of light in photonic crystals, enabling creation of diversified field pattern and energy flow landscape. 
\end{abstract}

\pacs{42.70.Qs, 78.20.Bh, 42.70.Df, 61.30.Jf}
\maketitle

\section{\label{sec:level1}Introduction}
Topological defects have been extensively studied in condensed matter physics. 
One notable example is screw/edge-dislocations in liquid crystals \cite{soft_matter}. 
It has been shown that the topological defects in liquid crystals can strongly influence the light-matter interaction \cite{marruci_PRL,Brasselet_PRL09,Brasselet_OL11,Barboza_PRL12,Brasselet_PRL12,Brasselet_PRL13,Barboza_PRL13,Loussert_APL14,Cancula_PRE14}. 
On one hand, the nematic disclinations may induce singularities in the light fields, generating optical beams with orbital angular momenta. 
On the other hand, strong light fields can imprint topological defects in liquid crystals, producing matter vortices \cite{Brasselet_JOPT10, Barboza_PRL12,Porenta_Soft12,Barboza_PRL13} . 
Such interactions, however, are not sufficient to confine light at the defects and create optical resonances, due to small refractive index variation in the liquid crystals. 

One efficient method of controlling light-matter interaction is to introduce a periodic modulation of the refractive index on the length scale of optical wavelength to make a photonic crystal \cite{joannopoulos, Soukoulis_book, Noda_book, Sakoda}.
By introducing structural defects to the photonic crystals, light may be tightly confined and forming resonances with high quality factor \cite{Painter}. 
Such defect cavities have also been used to manipulate light transport and generate optical vortices \cite{Sadreev,Saenz}. 
In addition, plasmonic nanostructures have been designed to mold the electromagnetic energy flow on the nanoscale via the formation and interaction of vortex nanogears \cite{Boriskina}. 
These studies illustrate different ways of producing diversified optical flow patterns for integrated photonic circuits. 

In this paper, we introduce topological defects to photonic crystals (PhCs). 
The unit cells are anisotropic and their orientations vary spatially to create topological defects in the orientational order. 
Compared to the liquid crystal, the interaction of light with the topological defect is greatly enhanced, as the unit cell is much larger than the liquid crystal molecule and the refractive index contrast is much higher. 
Instead of introducing disorder to the structural factor as the regular photonic crystal defects, we create disorder in the form factor. 
Despite that the topological defect breaks the translational symmetry of the photonic crystal, a residual photonic bandgap remains in the density of optical states. 
The band-edge states are dramatically modified in terms of spatial field profile and power flow pattern, e.g., the energy flux may circulate around the topological defect instead of flowing outwards. 
To further enhance optical confinement, we create the photonic crystal defect modes in these structures, and show that they are distinct from the regular defect modes, e.g. they may generate power flow vortices. 
Such characteristic features originate from the spatial variation of unit cell orientation. 

\section{\label{sec:level2} 2D square lattice with topological defect}

\begin{figure}[htbp]
\centering
\includegraphics[scale=0.29]{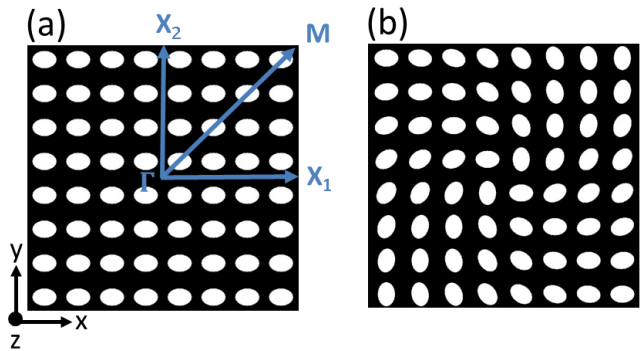}
\caption{ Introducing topological defect to a photonic crystal with anisotropic unit cell. (a) A square-lattice of air cylinders with elliptical cross-section embedded in a dielectric host. The major axis of the ellipses is parallel to the $\Gamma-X_1$ direction. (b) To introduce the topological defect, the major axis of each ellipse is rotated to an angle $\phi = k\theta + c$ from the $x$-axis. $\theta$ is the polar angle of the center position of the ellipse.  $k = 1$ and $c = \pi/4$. Away from the center, the ellipses in each of the four quadrant are aligned in the same direction, forming crystalline domains that rotate 90 degree from one quadrant to the next. }
\label{fig1}
\end{figure}

\begin{figure}[htbp]
	\centering
	\includegraphics[scale=0.19]{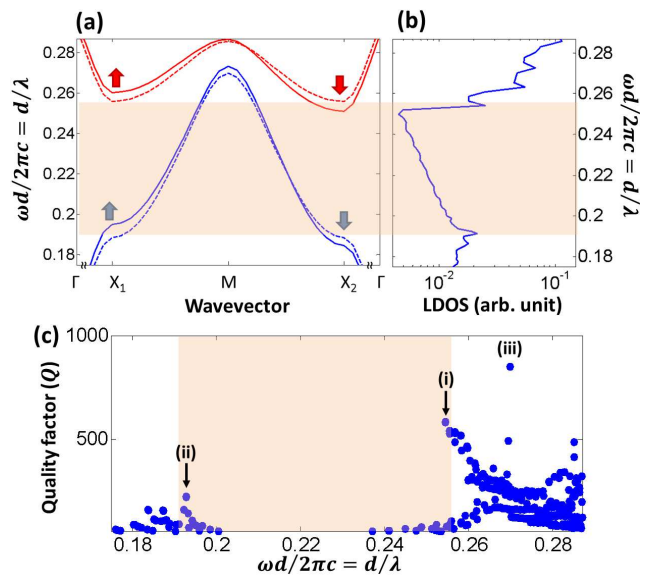}
	\caption{(Color online) Residual photonic bandgap (PBG) in the topological structure. (a) Dispersion relation of the first and second photonic bands with TE polarization in the square lattice of circular cylinders (dashed line) and elliptical cylinders (solid line). The refractive index of the cylinder is 1.0 and that of the dielectric host is 2.83. The circular cylinders have the radius $r = 0.3d$, where $d$ is the lattice constant. The elliptical cylinders occupy the same fraction of area, with aspect ratio $a/b = 1.4$. PBGs exist in the $\Gamma-X$ and $\Gamma-M$ directions, but the frequency degeneracies of the $\Gamma-X_1$ and $\Gamma-X_2$ bands are lifted once the cylinder cross-section deviates from circle.  (b) Local density of optical states at the center of the topological defect structure, exhibiting a reduction in the frequency range (marked by the shaded area) that corresponds to the PBG in the $\Gamma-X$ direction of the square lattice with circular cylinders. The residual PBG effect results from the long-range order in the elliptical cylinder position. (c) Quality-factor ($Q$) of TE-polarized modes in the topological defect structure with $32 \times 32$ cylinders. High-Q modes, e.g. the ones labeled (i)-(iii), are found close to the  band-edges of the corresponding periodic structure.}
	\label{fig2}
\end{figure}

\indent We start with a two-dimensional (2D) square lattice of air cylinders in a dielectric host. 
The lattice constant is $d$. The air cylinders have circular cross-section of radius $r$. 
The refractive index of the dielectric host is $n$. 
We calculate the photonic bands using the plane-wave expansion method \cite{mpb}. 
Figure \ref{fig2}(a) plots the first two bands of the transverse-electric (TE) polarization (electric field perpendicular to the cylinder axis) for $r/d$ = 0.3 and $n$ = 2.83. 
The photonic bandgaps (PBGs) are seen in the $\Gamma-X$ and $\Gamma-M$ directions, but they do not overlap to form a full (isotropic) gap at this filling fraction of dielectric material. 

Next we make the unit cell anisotropic by deforming the cross-section of each cylinder to an ellipse [Fig. \ref{fig1}(a)]. 
The major axis of the ellipse is $a$ and the minor axis is $b$. 
As the aspect ratio $a/b$ is varied, the cross-sectional area of the ellipse is kept constant and the structure has the same filling fraction of dielectric with the PhC of circular unit cell. 
The photonic bands are modified, as seen in Fig. \ref{fig2}(a) for $a/b$ = 1.4. 
The bands in the $\Gamma-X_1$ direction (parallel to the major axis of ellipse) are blue shifted, whereas the bands in the $\Gamma-X_2$ direction (parallel to the minor axis of ellipse) are red shifted. 
This can be understood from the shrinking (or expanding) of the dielectric region between two adjacent air cylinders in the $\Gamma-X_1$ (or $\Gamma-X_2$) direction, resulting in an decrease (or increase) of the effective refractive index for light propagating in the $\Gamma-X_1$ (or $\Gamma-X_2$) direction. 
Hence, the frequency degeneracy between the $\Gamma-X_1$ and $\Gamma-X_2$ band-edge modes is lifted by the deformation of the unit cell. 
The magnitude of their frequency splitting depends on the aspect ratio of the ellipses. 
The $\Gamma-X_1$ band-edge modes have electric field polarized in the $y$-direction ($E_y$), and the $\Gamma-X_2$ band-edge modes in the $x$-direction ($E_x$). 
Such a square-lattice photonic crystal with elliptical shaped scattering units has been used to generate polarized optical beams from photonic crystal surface emitting lasers \cite{noda_science} and as a polarization beam splitter \cite{shanhuifan_JOSAA}. 

Finally, we introduce the topological defect by rotating individual ellipses on the square lattice to different directions.
Following the configurations of wedge disclinations in liquid crystals, the angle between the major axis of an ellipse to the $x$-axis is set to $\phi = k \theta + c$, where $\theta$ denotes the polar angle of the center position of the ellipse,  $k$ is a half-integer or an integer that represents the topological charge, and $c$ is a constant. 
Fig. \ref{fig1}(b) shows such a structure with $k=1$ and $c=\pi/4$, which forms a structural vortex at the center. 
Note that the center of the topological defect, which coincides with the origin of the 2D coordinates, is chosen to be in the middle of four adjacent cylinders. 
Moving away from the center, the ellipses in each of the four quadrants are aligned in the same direction, albeit their direction rotate by 90 degree from one quadrant to the next. 
This structure is clearly non-periodic and does not possess any translational symmetry due to spatially varying orientation of the anisotropic unit cell.
Nevertheless, a residual photonic bandgap can be seen in the local density of optical states (LDOS) at the center of the structure in Fig. \ref{fig2}(b). 
The LDOS, $g(\textbf{r}, \omega) = ({2\omega}/ {\pi c^2}) \textnormal{Im}[G(\textbf{r},\textbf{r},\omega)]$, 
where $G(\textbf{r},\textbf{r'},\omega)$ is the Green's function for the propagation of the magnetic field $H_z$ from point $\textbf{r}$ to $\textbf{r'}$.  
is calculated numerically with a commercial program COMSOL \cite{comsol}.
The frequency range in which the LDOS is suppressed coincides with the PBG in the $\Gamma-X$ direction [Fig. \ref{fig2}(a)]. 
The residual PBG effect results from the long-range positional order of the cylinders in the structure. 
\section{\label{sec:level3} Optical resonances in photonic topological defect structure}
\indent In this section we calculate the optical resonances with high quality ($Q$) factor in the photonic topological defect structure of finite size. 
The structure consists of $N$ = 1024 air cylinders, forming a $32 \times 32$ square lattice. 
It is embedded in a dielectric host medium, which is surrounded by a perfectly matched layer to absorb light escaping from the structure. 
We calculate the TE polarized resonances using the finite-difference frequency-domain method \cite{comsol}. 
Due to light leakage, the resonant modes have complex frequencies $\omega = \omega_r - i \omega_i$, and the quality factor is $Q = \omega_r/2\omega_i$. 
Figure \ref{fig2}(c) plots the $Q$ factor versus the normalized frequency $\omega d / 2 \pi c = d / \lambda$ of resonances in the vicinity of the residual PBG shown in Fig. \ref{fig2}(b). 
High-$Q$ modes, e.g., the ones labeled (i)-(iii) in Fig. \ref{fig2}(c), correspond to the band-edge modes in the $\Gamma-X$ and $\Gamma-M$ directions of the square-lattice PhC. 
However, they are strongly modified by the topological defect, as will be seen below. 
\begin{figure}[htbp]
	\centering
	\includegraphics[scale=0.38]{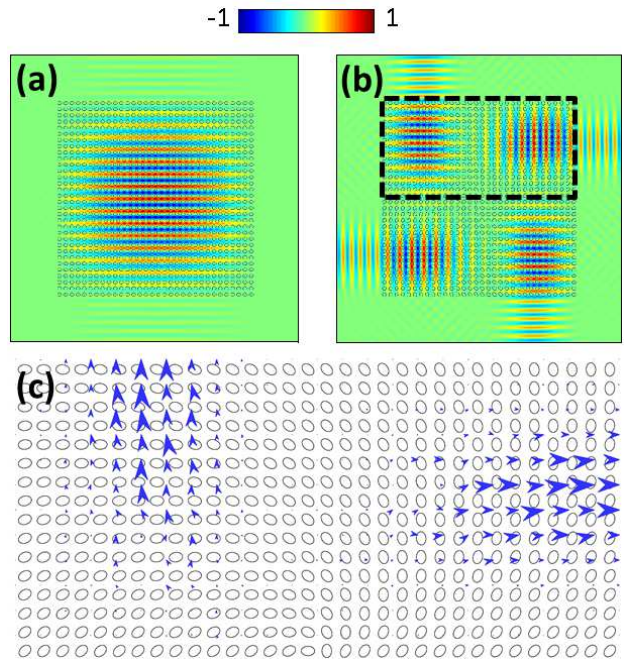}
	\caption{(Color online) Modification of $\Gamma-X_2$ air band-edge mode by topological defect. (a) Spatial distribution of the magnetic field ($\textnormal{Re}\{H_z\}$) for the TE-polarized air band-edge mode in the $\Gamma-X_2$ direction of a square-lattice with elliptical cylinders shown in Fig. \ref{fig1}(a). The structural parameters are the same as those in Fig. \ref{fig2} (a). Light propagates in the $y$-direction and escapes from the top and bottom boundaries of the lattice.  (b) The spatial distribution of magnetic field of a high-$Q$ mode, labeled (i) in Fig. \ref{fig2}(c), in the topological defect structure. The mode is split into four regions of high intensity, each of them  resembling the field pattern in (a) except a possible rotation. (c) The spatial map of the Poynting vector in the top half region highlighted in (b). Each arrow points in the direction of local energy flux, and its length is proportional to the amplitude of the flux. In each of the four domains, light escapes only in one direction, as the emission in the opposite direction is blocked by the adjacent domain due to local PBG effect. }
	\label{fig3}
\end{figure}

In Fig. \ref{fig3}(a,b), we compare the spatial field profiles of mode (i) in the topological defect structure and the corresponding band-edge mode in the PhC with elliptical cylinders [Fig. \ref{fig1}(a)]. 
The band-edge mode [Fig. \ref{fig3}(a)], located at the high frequency edge of the PBG in the $\Gamma-X_2$ direction, has maximum electric field intensity within the air cylinders, thus called an air band-edge mode. 
This mode is formed via distributed Bragg reflection of light propagating in the $y$-direction (parallel to the minor axis of the ellipse) by layers of air ellipses parallel to the $x$-axis, with the electric field polarized in the $x$-direction (parallel to the ellipse major axis). 
In the presence of topological defect, mode (i) is split into four parts, each occupying one quadrant [Fig. \ref{fig3}(b)]. 
The field profile in every quadrant resembles that of the air band-edge mode in Fig. \ref{fig3}(a), which rotates by 90 degree from one quadrant to the next. 
As mentioned earlier, the ellipses in each quadrant, especially the ones far from the center, are aligned in the same direction, thus forming crystalline domains that support PBG locally. 
Since the crystalline orientation rotates 90 degree from one quadrant to the next, the band-edge mode re-orient their spatial profiles. 
For the $\Gamma-X_2$ air band-edge mode, the direction of light propagation switches between $x$-axis and $y$-axis.
\begin{figure}[htbp]
	\centering
	\includegraphics[scale=0.38]{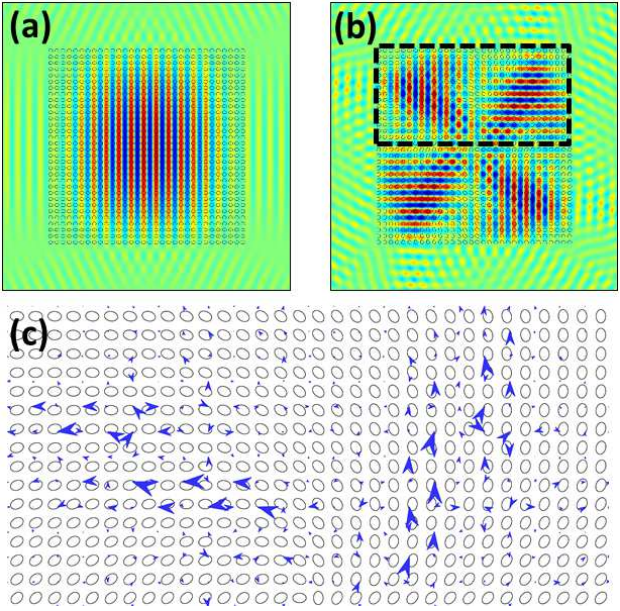}
	\caption{(Color online) Modification of $\Gamma-X_1$ dielectric band-edge mode by topological defect. (a) Spatial distribution of the magnetic field ($\textnormal{Re}\{H_z\}$) for the TE-polarized dielectric band-edge mode in the $\Gamma-X_1$ direction of a square-lattice with elliptical cylinders shown in Fig. \ref{fig1}(a). Light propagates in the $x$-direction and escapes from the left and right edges of the lattice.  (b) The spatial distribution of magnetic field of a high-$Q$ mode, labeled (ii) in Fig. \ref{fig2}(c), in the topological defect structure. The mode is split into four regions of high intensity, each of them resembles the field pattern in (a) except a possible rotation. (c) The spatial map of the Poynting vector in the top half region of (b). Each arrow points in the direction of local energy flux, and its length is proportional to the amplitude of the flux. In each of the four domains, light escapes only in one direction, as the emission in the opposite direction is blocked by the adjacent domain due to local PBG effect.
	 }
	\label{fig4}
\end{figure}

Due to the rotation of crystalline domains, light escapes from the topological defect structure asymmetrically. 
To visualize the energy flow, we calculate the Poynting vector averaged over one optical cycle,  $\vec{S}(x,y) = \frac{1}{2}\textnormal{Re}[\vec{E}(x,y)\times \vec{H}^*(x,y)$]. 
Figure \ref{fig3}(c) is a spatial map of the Poynting vector in the top half of the structure [highlighted in Fig. \ref{fig3}(b)]. 
Each arrow in the map points in the direction of local energy flux, and its length is proportional to the amplitude of the flux. 
In the top right quadrant the energy flows rightward, while in the top left quadrant the energy flows upward. 
The change in the flow direction follows the switch in the crystalline orientation. 
In each quadrant the energy flow is not symmetric, e.g., in the top right quadrant light escapes to the right but not to the left.  
This is because the crystalline domain on the left, which is rotated by 90 degree, blocks the light. 
Once leaving the top right domain and entering the top left one, the emission from the $\Gamma-X_2$ air band-edge mode begins propagating in the $\Gamma-X_1$ direction. 
Since $\Gamma-X_1$ air band-edge has higher frequency than the $\Gamma-X_2$ air band-edge, the $\Gamma-X_2$ air band-edge mode falls into the PBG in the $\Gamma-X_1$ direction, and its output is reflected by the left domain.  

\begin{figure}[htbp]
\centering
\includegraphics[scale=0.375]{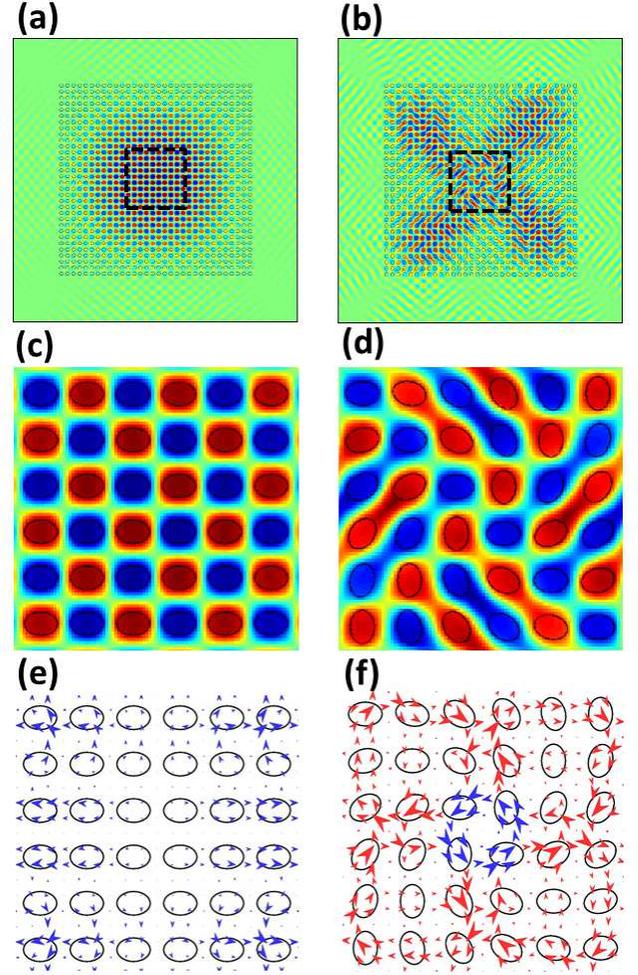}
\caption{(Color online) Modification of $\Gamma-M$ dielectric band-edge mode by topological defect. (a) Spatial distribution of the magnetic field ($\textnormal{Re}\{H_z\}$) for the TE-polarized dielectric band-edge mode in the $\Gamma-M$ direction of a PhC with elliptical cylinders shown in Fig. \ref{fig1}(a). The structural parameters are the same as those in Fig. \ref{fig2} (a). (b) The spatial distribution of magnetic field of a high-$Q$ mode, labeled (iii) in Fig. \ref{fig2}(c), in the topological defect structure. The mode becomes distorted, forming a cross pattern. (c,d) Zooming into the central region of the field patterns in (a,b), revealing the distortion caused by the topological defect. 
(e,f) The spatial map of the Poynting vector for the modes shown in (a,b). Each arrow points in the direction of local energy flux, and its length is proportional to the amplitude of the flux. In the PhC with elliptical cylinders, the energy flows outwards to the boundary. In the topological defect structure, the energy flux circulates around the defect center (colored in blue), forming an optical vortex. There are additional four circulating flows in the four quadrants.  }
\label{fig5}
\end{figure}
In contrast, the $\Gamma-X_1$ air band-edge mode in each domain can leak to the adjacent one, because its frequency falls outside the PBG in the $\Gamma-X_2$. 
Consequently, its $Q$ factor is much lower. 
Similar behavior is expected for the $\Gamma-X_2$ dielectric band-edge mode in the topological defect structure whose frequency is beyond the PBG of $\Gamma-X_1$. 
However, the $\Gamma-X_1$ dielectric band-edge mode is different because its frequency falls into the PBG of $\Gamma-X_2$, and it contributes to the high-$Q$ mode (ii) in Fig. \ref{fig2}(c). 
In the PhC with elliptical cylinders, the $\Gamma-X_1$ dielectric band-edge mode propagates in the $x$-direction (parallel to the major axis of the ellipse), with the electric field polarized in the $y$-direction (parallel to the minor axis of the ellipse) [Fig. \ref{fig4}(a)]. 
Since the $\Gamma-X_1$ dielectric band-edge has higher frequency than the $\Gamma-X_2$ dielectric band-edge, the $\Gamma-X_1$ dielectric band-edge mode in each domain of the topological structure cannot penetrate into the adjacent domain which is rotated by 90 degree [Fig. \ref{fig4}(b)]. 
Consequently, the energy flows only outwards, as shown in Fig. \ref{fig4}(c).  
Moreover, the topological defect introduces structural disorder in each domain, especially close to the center, hence the  $\Gamma-X_1$ dielectric band-edge mode becomes coupled to nearby low-$Q$ resonance, causing a distortion of its spatial field profile. 

In addition to the $\Gamma-X$ band-edge modes, we also observe high-$Q$ modes near the $\Gamma-M$ band-edges. 
One example is mode (iii) in Fig. \ref{fig2}(c), which is located close to the dielectric band-edge.  
A comparison of its field profile to that of the $\Gamma-M$ band-edge mode in the square-lattice PhC reveals a strong effect from the topological defect [Figure \ref{fig5}(a,b)].
Mode (iii) is split into five regions of high intensity, in addition to the four quadrants, there is an additional one at the structure center. 
This is different from  modes (i) and (ii) which have diminishing intensity at the center. 
Zooming into the field profile at the center, the dielectric band-edge mode in Fig. \ref{fig5}(c) displays a checkerboard pattern with the wavevector in the $\Gamma-M$ direction. 
The continuous rotation of ellipses around the center of the topological defect structure causes a significant distortion of the field pattern, as seen in Fig. \ref{fig5}(d).  
In Fig. \ref{fig5}(e), the spatial map of the Poynting vector in the square-lattice PhC shows that the energy flowing out of the structure in all four directions. 
In the topological defect structure, however, the energy flux circulates inside, forming an optical vortex around the defect center [Fig. \ref{fig5}(f)]. 
Besides the counter-clockwise (CCW) circulating flow at the center, there are additional clockwise (CW) circulating flows in the four quadrants. 
They result from the spatial variation of the ellipse orientation, namely, the spatial inhomogeneity of the form factor.   

\section{Strongly-confined defect states}
\begin{figure}[htbp]
\centering
\includegraphics[scale=0.25]{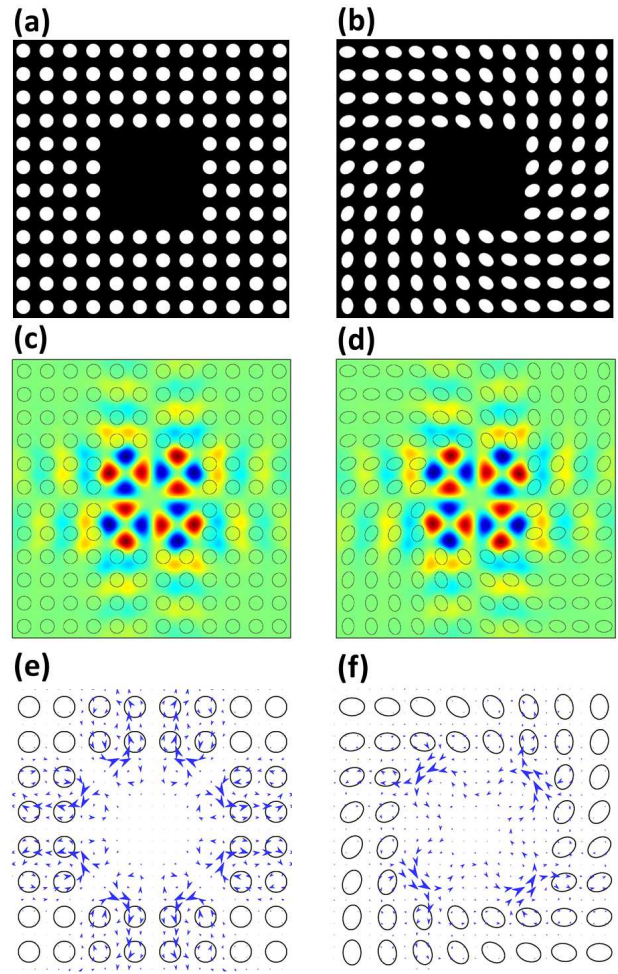}
\caption{(Color online) Localized defect state in the topological defect structure, in comparison to the PhC defect state. (a) Removing $4 \times 4$ cylinders from the center of a square lattice of circular cylinders to create photonic defect states. (b) Removing $4 \times 4$ cylinders from the center of the topological defect structure. The structural parameters remains the same as in previous figures. (c,d) The spatial distribution of the magnetic field for the highest-$Q$ defect state in (a,b). The field profile is nearly unchanged by the topological defect. (e,f) The spatial map of the Poynting vector for the defect states in (c,d). Each arrow points in the direction of local energy flux, and its length is proportional to the amplitude of the flux. In (e), the energy flows out of the central defect region in all four directions. In (f), the optical flux circulates counter clockwise (CCW) in the central region. }
\label{fig6}
\end{figure}

In the previous section, we have shown that by introducing topological defect into a square-lattice of elliptical cylinders, the band-edge modes are significantly modified. 
Although their $Q$ factors are the highest among all the resonances, such modes remain spatially extended and light can easily escape through the boundaries. 
To tightly confine light within the topological defect structure, we remove $4 \times 4$ ellipses from the center to create localized states similar to the photonic defect states of a regular PhC [Fig. \ref{fig6}(a,b)]. 
Among all the defect states that are generated within the residual PBG, the one shown in Fig. \ref{fig6}(d) has the highest $Q$ of $1.27 \times 10^4$, which is more than one order of magnitude higher than the rest. 
Its field profile, presented in Fig. \ref{fig6}(d), resembles that of a regular photonic crystal defect state shown in Fig. \ref{fig6}(c). 
However, the energy flow pattern becomes notably different, as seen in Fig. \ref{fig6}(e,f). 
For the regular photonic crystal defect state, the energy flows out of the central defect region [Fig. \ref{fig6}(e)]. 
In the presence of topological defect, the optical flux circulates CCW in the central region [Fig. \ref{fig6}(f)]. 

\begin{figure}[htbp]
\centering
\includegraphics[scale=0.25]{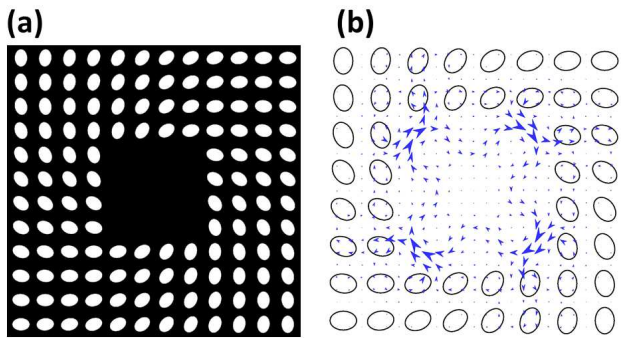}
\caption{(Color online) Switching the direction of energy flow in the defect state in Fig. \ref{fig6}. (a) The topological defect structure with the major and minor axes of the ellipse exchanged, which is equivalent to rotate all the ellipses in the structure shown in Fig. \ref{fig6}(b) by 90 degree. (b) The spatial map of the Poynting vector for the defect state in (a), showing the power circulation flips the direction from CCW to CW. }
\label{fig7}
\end{figure}

\indent The drastic change in the energy flow can be explained by the orientation of the ellipses along the boundary of the defect region. 
In Fig. \ref{fig6}(a), with circular cylinders, the waves propagating in the CW and CCW directions in the defect region experiences the same amount of scattering from the cylinders on the boundary. 
In Fig. \ref{fig6}(b), the orientation of the ellipses keeps changing along the boundary of the defect region, thus the CW and CCW waves experience different amounts of scattering, breaking the balance in amplitude between the CW and CCW waves. 
For the high-$Q$ defect state, the CW wave experiences more scattering loss than the CCW wave, thus the net energy flow inside is CCW. 
To confirm this mechanism, we switch the major and minor axes of the ellipse, which is equivalent to rotate all the ellipses by 90 degree [Fig. \ref{fig7}(a)]. 
In this structure, the CCW wave experiences more scattering loss than the CW wave, thus the high-$Q$ defect state is dominated by the CW wave, as confirmed from the flux pattern in Fig. \ref{fig7}(b).   \\

\section{Conclusion}

In summary, we introduce topological defect into a square-lattice of elliptical cylinders. 
Although the translational symmetry is broken, the residual photonic bandgap effect is evident from the suppressed density of optical states, which
results from the long-range positional order of the ellipses. 
The band-edge modes are strongly modified by the spatial variation of the ellipse orientation, both the field profile and energy flow are altered. 
For example, the high-$Q$ mode located at the edge of $\Gamma-X$ bandgap splits into four regions of high intensity and the output flux becomes asymmetric due to spatially separated and rotated crystalline domains.  
The $\Gamma-M$ band-edge mode is also modified significantly and the energy flux circulates around the topological defect center. 

To enhance the optical confinement, we remove the elliptical cylinders at the center to create localized states, in analogy to the photonic crystal defect states.   
Such states are dominated by either CW or CCW circulating waves, due to unbalanced scattering loss from the elliptical cylinders at the boundary of the defect region. 
The flow direction can be switched by changing the ellipse orientation. 
The intrinsic optical vortices may be useful for modifying the motion of small particles or enhancing the interaction with chiral molecules for sensing applications. 

Although this study is focused on a specific type of topological defect, we believe the photonic crystals with various types of topological defects can support diversified resonances with distinct characteristic.
The deterministic aperiodic variation of the unit cell orientation adds another dimension to the control of light in a photonic crystal, enabling creation of desired field pattern and energy flow landscape. 

\section{Acknowledgement}
We thank Yaron Bromberg, Eric Dufresne and Chinedum Osuji for useful discussions. This work is supported by the MURI grant No. N00014-13-1-0649 from the US Office of Naval Research. \\

\end{document}